\documentclass[manuscript]{acmart}

\usepackage{booktabs, siunitx, hyphenat} 
\usepackage[raggedrightboxes]{ragged2e}

\AtBeginDocument{%
  }

\setcopyright{acmcopyright}
\copyrightyear{2024}
\acmYear{2024}
\acmDOI{10.1145/3613905.3644060}

\acmPrice{00.00}
\acmISBN{xxxx}

\begin{document}

\title{Toward Humanity-Centered Design without Hubris}

\author{Tim Gorichanaz}
\email{gorichanaz@drexel.edu}
\orcid{0000-0003-0226-7799}
\affiliation{%
  \institution{Drexel University}
  \streetaddress{3675 Market Street}
  \city{Philadelphia}
  \state{Pennsylvania}
  \country{USA}
  \postcode{19104}
}

\begin{abstract}
Humanity-centered design is a concept of emerging interest in HCI, one motivated by the limitations of human-centered design. As discussed to date, humanity-centered design is compatible with but goes beyond human-centered design in that it considers entire ecosystems and populations over the long term and centers participatory design. Though the intentions of humanity-centered design are laudable, current articulations of humanity-centered design are incoherent in a number of ways, leading to questions of how exactly it can or should be implemented. In this article, I delineate four ways in which humanity-centered design is incoherent---which can be boiled down to a tendency toward hubris---and propose a more fruitful way forward, a humble approach to humanity-centered design. Rather than a contradiction in terms, ``humility'' here refers to an organic, piecemeal, patterns-based approach to design that will be good for our being on this earth.
\end{abstract}

\begin{CCSXML}
<ccs2012>
   <concept>
       <concept_id>10003120.10003123.10011758</concept_id>
       <concept_desc>Human-centered computing~Interaction design theory, concepts and paradigms</concept_desc>
       <concept_significance>500</concept_significance>
       </concept>
   <concept>
       <concept_id>10003120.10003121.10003126</concept_id>
       <concept_desc>Human-centered computing~HCI theory, concepts and models</concept_desc>
       <concept_significance>300</concept_significance>
       </concept>
 </ccs2012>
\end{CCSXML}

\ccsdesc[500]{Human-centered computing~Interaction design theory, concepts and paradigms}
\ccsdesc[300]{Human-centered computing~HCI theory, concepts and models}

\keywords{human-centered design, humanity-centered design, Don Norman, critique, Christopher Alexander}


\maketitle

\section{Introduction}

``Saving humanity is in vogue right now,'' remarks an \emph{Economist} article from December 2023 \citep{Economist2023}. The piece explores the messiah complex of tech-sector mogul Elon Musk and compares it to the hubris of other innovators, from Sam Bankman-Fried, the founder of cryptocurrency exchange FTX who was also recently convicted of fraud, to Henry Ford, the founder of Ford Motor Company who was also an antisemitic conspiracy theorist.

Musk has become a poster boy for world-saving ambitions, from his aspirations to make humans an interplanetary species to ushering in the sustainable energy revolution \citep{Isaacson2023}, but he is not alone. The past year has brought renewed hype regarding the existential threats of artificial intelligence, as well as fears engendered by growing political unrest and international conflicts the world over. Not to mention the continual, looming threat of climate change \citep{Light2017}. Indeed, perhaps big thinking is the necessary response to our big problems. 

In the world of design, such big thinking has has become manifest in the emerging concept of humanity-centered design, as distinct from human-centered design. At first blush, it seems that humanity-centered design presents a welcome expansion of the circle of concern for designers and a stronger inbuilt ethical orientation for design work. However, it risks falling prey to the same sort of messiah complex that is visible in public figures such as Elon Musk and his predecessors. That is, it risks hubris.

In this article, I argue for an alternative, more humble, vision for humanity-centered design, one that relies on organic, incremental growth rather than the long-term mapping of revolutions. It may seem less grandiose in the short term, but in the end it will be more effective and humane. This vision is based on the philosophy of architect and design theorist Christopher Alexander, particularly as expressed in his master plan for the University of Oregon's campus \citep{Alexander1975}. Alexander is best known in HCI for his work on design patterns \citep{Wania2015}, but his larger project was certainly resonant with the goals of humanity-centered design \citep{Gorichanaz2023alexander}. 

Before discussing Alexander's vision, though, I will review the emergence and criticisms of human-centered design and outline four ways in which the current expression of humanity-centered design is incoherent. In the context of this paper, I suggest that Alexander's philosophy presents not only a more effective but also a more coherent vision for humanity-centered design.  

\section{Human-Centered Design and Its Limitations}

At the dawn of computer programming, there wasn't much thought given to ease of use. Computers were used by highly-trained experts, often the people who built the machines themselves. But by the 1960s, computer manufacturers, programmers and users were increasingly different people, and soon the field of HCI emerged, dedicated to making computing systems more usable---more effective, efficient, learnable and so on. 

In its early years, HCI was narrowly focused on these issues of usability engineering \citep{Grudin1990}. But it wasn't long before critics rallied for a broader set of concerns, paving the way for the new paradigm of human-centered design. A landmark paper in this regard was Gould and Lewis' 1985 articulation of three principles to ensure usability: an early focus on users and tasks, empirical measurement, and iterative improvement \citep{GouldLewis1985}. Later, Don Norman summarized the key tenets of human-centered design as: addressing the root issues rather than symptoms, focusing on people's needs and capabilities, taking a systems perspective, and continually testing and refining proposed solutions \citep{Norman2013}. More broadly, human-centered design provides a process and methods to address human contexts and needs, leading to more efficiency, productivity and positive experiences \citep{Sharp2023}. 

Human-centered design undoubtedly enables usability and high-quality user experiences, but in recent years it has become clear that the methods of human-centered design is not enough to reliably lead to products that meet human needs, particularly in a moral sense. 

First, witness that even though human-centered design predominates in design education (particularly in HCI and UX), it is not always carried out in practice. Consider how throughout 2023, countless tech firms scrambled to infuse their products and services with generative artificial intelligence (AI). In January 2023, for example, Microsoft announced plans to add AI capabilities into virtually all its products \citep{Sinclair2023}, and Google followed after \citep{Love2023}. In the first half of 2023, generative AI startups collectively raised \$14.1 billion in disclosed equity funding, which was more than in all of 2022 and despite an overall contraction in the investment climate \citep{CBI2023}. These are examples not of human-centered design but of technology-driven design. 

But even when human-centered design is implemented, the results do not always shine. Early critics observed that human-centered design may be superficially or incorrectly applied, for instance by focusing on narrowly-defined user tasks rather than more contextualized human activities \citep{Gasson2003,Norman2005,Bannon2011}. More recently, critics have raised other points: human-centered design creates products for the marketplace, serving short-term desires rather than long-term human futures, meaning or connection \citep{chapman2021,DunneRaby2013}; in human-centered design, innovation itself is considered progress, disincentivizing systemic change \citep{Harris2021}; and human-centered design is overly anthropocentric, not attending to the needs of our planet and other species we live with \citep{Wakkary2021}.

Efforts to overcome these sorts of issues gave rise to what has been called the ``third wave'' of HCI \citep{Bodker2006,Harrison2011}. In this third wave, HCI has become more concerned with the contextual, social, affective and emergent dimensions of user experience \citep{Bodker2006,Gaver2022,Harrison2011}. Moreover, deeper questions such as those of worth, identity and meaning have risen to the surface \citep{Cockton2006,Cockton2020,Fallman2011,Kaptelinin2018,Mekler2019}. Frameworks such as participatory design and value sensitive design began to mature, and naturalistic research became more common \citep{Rogers2017}. 

\section{The Emergence of Humanity-Centered Design}

Among third-wave HCI researchers and designers---as well as increasingly among the public---it is being recognized that technology is not morally neutral and that ignoring the moral component inherent in design is one of the reasons negative consequences result \citep{ConsilienceProject2022}.

Consonantly, a new orientation toward human-centered design has been emerging, sometimes termed ``humanity-centered design,'' which specifically foregrounds moral issues \citep{Gonsher2017,Norman2023,Reed2005,Sklar2010}. And, to be sure, there are also discussions of design and moral issues that do not refer to this term, such as design justice \citep{CostanzaChock2020,Tunstall2023} and value sensitive design \citep{Friedman2019}. Notably, the term ``humanity-centered'' is beginning to find footing in industry as well, with examples including the UX consulting firm HmntyCntrd and the advocacy organization the Center for Humane Technology. 

While the concept of humanity-centered design is still emergent (perhaps it is currently where human-centered design was in 1985), its key principles are beginning to cohere. In his 2023 book \emph{Design for a Better World}, Don Norman \citep{Norman2023} articulates five key principles for humanity-centered design, which he describes as compatible with but going beyond human-centered design: addressing the root problem, focusing on the ecosystem, taking a long-term systems perspective, continually testing and refining, and designing with the community as much as possible. A side-by-side comparison of Norman's principles for human-centered and humanity-centered design is presented in Table~\ref{table:norman}.

\begin{table}[bt]
\begin{tabular}{ l p{2.5in} p{2.5in} }
  & Human-Centered Design                                                                                                             & Humanity-Centered Design                                                                                                                                                                                                                                       \\
  \hline
1 & Solve the core, root issues, not just the problem as presented (which is often the symptom, not the cause).                       & Solve the core, root issues, not just the problem as presented (which is often the symptom, not the cause).                                                                                                                                                    \\
2 & Focus on the people.                                                                                                              & Focus on the entire ecosystem of people, all living things, and the physical environment.                                                                                                                                                                      \\
3 & Take a systems point of view, realizing that most complications result from the interdependencies of the multiple parts.          & Take a long-term, systems point of view, realizing that most complications result from the interdependencies of the multiple parts and that many of the most damaging impacts on society and the ecosystem reveal themselves only years or even decades later. \\
4 & Continually test and refine the proposed designs to ensure they truly meet the concerns of the people for whom they are intended. & Continually test and refine the proposed designs to ensure they truly meet the concerns of the people and ecosystems for whom they are intended.                                                                                                               \\
5 & --                                                                                                                                & Design with the community and as much as possible support designs by the community. Professional designers should serve as enablers, facilitators, and resources, aiding community members to meet their concerns.                                            
\end{tabular}
\caption{Side-by-side comparison of principles of human-centered and humanity-centered design, as articulated by Don Norman \citep[pp.~182--183][]{Norman2023}.}
\label{table:norman}
\end{table}

\section{Paradoxes in Humanity-Centered Design}

Prima facie, the vision of humanity-centered design may seem agreeable and straightforward. But more deeply, it harbors several paradoxes that may not be soluble. One might be tempted to embrace the incoherent vision of humanity-centered design for its rhetorical power more than for its logical validity. But on the other hand, a more coherent vision for the concept would likely stand up better to future developments and be more useful as an empirical guide for design and research. Since the concept of humanity-centered design is still emergent, it would be worthwhile to explore and attempt to resolve these paradoxes.

\subsection{Universalism vs. Localism}

The term ``humanity'' can be understood in two senses: the first meaning the entire human race (i.e.,~``humankind''), and the second referring to a virtue of love and benevolence (as in ``humane''). 

While some authors in the past have championed humane interfaces (``humanity'' in the latter sense) \citep{Cousins1989,Raskin2000}, writers explicitly invoking ``humanity-centered design'' primarily seem to interpret ``humanity'' in the former sense. For example, Norman writes that humanity-centered design involves ``the societal level of world populations'' \citep{Norman2022} and Kr\"uger et al. \citep{Krueger2023} write that ``it is important to consider all of humanity~\ldots~when designing.'' (Not only that, but these authors also maintain that humanity-centered design also considers ecosystems and other creatures besides humans, despite the term's reference only to humans.)

Using ``humanity'' in the first sense suggests universalism: the idea that a single design solution could work for all of humanity. This notion has been regarded as ultimately impossible and unworkable (if utopian) in discussions of universal design---even if, as a rhetorical device, the term ``universal design'' has been helpful for championing design decisions that include people who were previously excluded by designed systems \citep{Godden2016}. Universalism in design has been even called harmful, as it erases the meaningful distinctions across certain groups \citep{CostanzaChock2020}. There is also the practical difficulty of doing participatory design---which Norman suggests is inherent to humanity-centered design---with the entire world population, which I will also address below.

In discussions of human-centered design, particularly design thinking, the consensus seems to be that local solutions are preferable to universal ones, as inevitably people in different places have different needs and preferences (due to climate, history, language, etc.) \citep{Brown2010}. Incidentally, elsewhere in Norman's recent statement on humanity-centered design \citep{Norman2023}, he calls for local solutions rather than universal ones. In the end, the call for both universalism and localism may be an impossible circle to square. 

\subsection{Revolution vs. Incrementalism}

The very term ``humanity-centered'' evokes the big---note the allness of the word ``humanity'' compared to the smallness of ``human.'' Commensurately, humanity-centered design is wrapped up in discussions of righting global-scale historical harms \citep{Tunstall2023}, confronting existential crises \citep{Light2017}, upending global capitalism \citep{chapman2021} and the like, often invoking terms like ``transformation,'' ``disruption'' and ``dismantling.'' These discussions imply sweeping changes. And indeed, regarding the practicalities of funding projects, Norman observes that it is often easier to get funding for large projects than small ones, as institutional structures favor providing large sums at once rather than small, incremental ones \citep{Norman2023}. (Christopher Alexander observes something similar when it comes to projects on university campuses---there are institutional channels to fund the multi-million-dollar new building for this or that, but smaller improvements and maintenance are much harder to get done \citep{Alexander1975}.)

Yet at the same time, Norman argues that incremental steps are preferable to revolutionary, disruptive ones \citep{Norman2023}. As Norman writes, small steps allow a clearer signal in testing and refinement, they are more responsive to the inevitable change of the real world (which will be ongoing while a large project is underway), and they allow people and cultures to adapt to new systems and ways of being. 

Unlike the previous paradox, this one may be possible to resolve: ``We can address large issues through a multitude of small, flexible projects,'' Norman writes \citep[p.~54]{Norman2023}. 

\subsection{Longtermism vs. the Present}

Whereas human-centered design focuses on the near-term and present---seeing as it addresses people's needs today, often through industrial market structures that assess profitability on a quarterly basis---humanity-centered design is said to be long-term. 

Judging from authors such as Norman on humanity-centered design, ``long-term'' here is meant primarily to evoke environmental sustainability for addressing climate change. But more generally, orienting our efforts with respect to the long term falls under the rubric of the philosophical movement of ``longtermism,'' which argues that positively influencing humanity's long-term future is a moral obligation of our time. One line of reasoning goes, as MacAskill puts it, that if \emph{Homo sapiens} endures for the average mammalian species lifespan of one million years, then well over 99 percent of total humans will live in the future, and so at our current time of tremendous technological efflorescence we have the opportunity to improve their lives---or worsen them \citep{MacAskill2022}. 

When it comes to design, the fear is that in our efforts today will become locked in and difficult to change. Consider the lock-in evident in the form-factor of most smartphones (a single slab without a tactile keyboard, all of similar sizes) or the architecture of most social media websites (news feed, profile, messaging). Further, designs embed human values, which may then lead to the lock-in of suboptimal values for the long-term future \citep{MacAskill2022}. All this raises the stakes for and motivates today's designers to consider the long term in their designs, perhaps even the very long term. 

Certainly, again considering the existential threats that humanity faces \citep{Light2017}, today's designers could be said to have certain obligations with respect to the long term. However, we must also remember that there are near-term issues that must be addressed. Future people are hypothetical, but people alive today are real. Even if there will be more future people than present ones, their suffering and problems are still only hypothetical, and those future people may never arrive if present suffering is not addressed. 

There are numerous other issues with longtermism, such as the assumption that technological progress is good and that ``stagnation'' is bad, the implication that the goal of humanity should be to lock in ``good'' values if possible rather than avoid lock-in altogether, and the notion that one single index for human well-being can be used in a valid way for design and policy. Further, discussions of longtermism seem to imply universalism in the sense of global cooperation and one-size-fits-all solutions, thus falling prey to the issues of universalism mentioned above. Operationalizing longtermism also faces issues with respect to participation, which I will discuss below (since they also apply to the present and near-term).

A full critique of longtermism is outside the scope of this article, but there are several such published works, in both academe \citep{Adams2023,Tarsney2023} and the public sphere \citep{Torres2022,Samuel2022}.

In the end, the meat of longtermism may be more philosophical puzzle-play than anything. In MacAskill's 2022 book, which serves as the most comprehensive statement on longtermism to date, the recommendations he arrives at are things like considering vegetarianism, choosing a career that is meaningful and does some good for the world, and doing activism---conclusions that can be easily reached without longtermism. This suggests that even the strongest adherents to longtermism cannot find any implications of the philosophy over and above what is already implied by other, less problematic, philosophies.

\subsection{The Impossibility of Participation}

A final paradox of humanity-centered design is its impossible call for participation. To be sure, participatory design has been developed for decades \citep{Floyd1989,Mumford1979,Schuler1993}---arising along with the origins of human-centered design. And indeed, the suite of participatory approaches have come to be seen as approaches under the umbrella of human-centered design \citep{Sharp2023}. The general idea is that people should have some agential stake in the designed products and services they will be using. 

Humanity-centered design attempts to take this further. As articulated by authors such as Norman \citep{Norman2023}, humanity-centered design calls for entire populations and ecosystems to be participants in design---both in the present and in the (long-term) future. That is, animals, plants, all present and future humans, and presumably other organisms and natural systems. 

Yet how these are all to be included is left unstated. Most of these entities cannot represent their own interests and needs to human designers without human advocates who must interpret those interests and needs, probably incorrectly. Of course, guesses can be made---and to be sure all living beings share at least certain needs, such as that of nourishment---but beyond the very general it is difficult to identify universal needs and preferences that are not debatable. The more unlike us and the farther in time, the more difficult it is. 

Even if all these entities could speak in human language, the practical reality remains of adjudicating billions and billions of viewpoints to arrive at a single design solution. It is simply intractable. (And considering fewer participants in a participatory approach returns us to human-centered design as already understood.)

\section{The Hubris of Humanity-Centered Design}

The above paradoxes of humanity-centered design can be boiled down to a single axis: that of arrogance and humility. (Strictly speaking, humility can be said to sit in the middle of a continuum between arrogance on one end and diffidence on the other \citep{Church2017}; but in the discussions of humanity-centered design to date, diffidence has not been an issue.)

In individual terms, humility is the virtue of having an accurate view of oneself, including one's strengths and weaknesses, and being responsive to those in others \citep{VanTongeren2019,Wright2019}. This involves a lack of concern for self-importance and the acceptance of things that are out of our control. Humility is about doing what one can with the resources one has, not biting off more than one can or should chew. Arrogance, in contrast, is having an overinflated view of oneself, such as presuming to have knowledge that one does not have and endeavoring to influence things that are not in one's control. 

In philosophy and psychology---and increasingly in the everyday world of business and other arenas \citep{Grant2021}---humility is recognized as a positive and prosocial character skill that encourages positive affect and growth, while arrogance is recognized as a defect that inhibits positive affect and growth \citep{Wright2019}. Hubris, going one step further than arrogance, is arrogance whose ultimate vice has been borne out. It typically refers to the failure resulting from excessive arrogance, such as in Greek tragedy.

Though humility and arrogance are generally descriptors applied to a person, they can also characterize organizations and movements in analogous ways. In my view, the conceptualizations of humanity-centered design and human-centered design straddle the axis of humility and arrogance. On one hand, seemingly inherent to humanity-centered design, are examples of messiah-complex hubris such as the case of Sam Bankman-Fried, whose grandiose visions for the future of humanity led him astray \citep{Economist2023}. And on the other hand is humility, closer to human-centered design as currently practiced, which seeks to take small steps and slowly refine them, moving toward preferred solutions without overstepping.

But this would be a troublesome conclusion, given the already well-documented limitations of human-centered design mentioned above. Despite the paradoxes, humanity-centered design is attempting to address valid and urgent concerns. Is it possible to envision a humanity-centered design without the hubris? 

\section{A Model for Humanity-Centered Design without Hubris: The Case of a University Master Plan}

What do designers do? They design. What do they design? They design designs. Right in our language is the notion that designers essentially draw a map of the future, which then gets built. Even if the reality is messier, that seems to be the implicit ideal. Designers do their work and then hand it off to the people who will build it. Though it is recognized that the designer could or should also be involved during development \citep{Buxton2007,Feng2023}, the goal of that involvement seems to be ensuring that the map is implemented correctly. I call this the map-of-the-future conceptualization of design. 

Map-of-the-future design surely has its limitations. Indeed, handoff remains a major barrier to human-centered design practice in the workplace \citep{Feng2023}. And if handoff is such an issue when designs are to be developed right away, consider how much more of a challenge it would be when the design is meant to be built over years or decades, or when it is meant to last millennia. 

If there is to be humanity-centered design without hubris, an alternative to the map of the future must be possible. I believe the seeds of a solution can be found in a discussion of university master plans dating back to the 1970s. 

\subsection{The Map-of-the-Future Approach and Its Limitations}

University master plans are maps of the future par excellence. At their best, they are created with community involvement and iteration, yet in the end they provide a map for the next several years of university development. 

Let us consider my own university's Master Plan as an example \citep{Drexel2012}. Published in 2012, the latest date to be found in the vision document \citep{DrexelVision2012} is 2017, suggesting the plan represented a five-year guide. (Though, confusingly, the webpage for the plan reads, ``The Master Plan's timeframe\ldots{} outlin{[}es{]} 30 years of campus growth and development.'')

Looking at this Master Plan today, it is immediately evident that some of the projects mentioned in the document came to fruition while others did not. The ``plan'' gives no indication about why (e.g., prioritization, contingencies). Further, several projects came about that do not appear in the plan at all, such as my own college's creation in 2013 \citep{CCIAbout} and its relocation to space in an office building that opened in 2019 to accommodate its growth \citep{CCIMove}, as well as a life sciences building whose construction is currently underway \citep{DrexelLifeSci2022}. 

The point is not that this particular Master Plan was uniquely bad, but that creating the plan as a map of the future was already a doomed strategy. A map is simply too rigid for the task. That's because, if anything changes in the intervening years, such as the formation of a new college, a map of the future gives no indication for how to adapt.

For example, my university's Master Plan includes a call to ``emphasize close relationships and short travel times between related programs to encourage cross-disciplinary collaboration.'' But how could that be squared with the reality of a newly formed and fast-growing college that quickly outgrows its space and then out of exigency must relocate to a building off campus? So much for short travel times; many of my students now arrive late to class because it is physically impossible to make it across campus and up the elevator within the scheduled 10 minutes between classes---and God forbid they have to use the bathroom.

\subsection{A Better Approach}

So if the map-of-the-future approach doesn't work for planning a vibrant university, what might be better?

This was essentially the question asked by architect Christopher Alexander (who is best known in computing for originating the concept of design patterns \citep{Wania2015}) when he led the development of the University of Oregon's master plan in the early 1970s. In that project, Alexander and his colleagues created a master plan that was not a map of the future but rather a philosophy and set of patterns for how the campus should adapt over time. As Alexander describes the project in his book \emph{The Oregon Experiment} \citep{Alexander1975}, the major principles include:

\begin{description} 
    \item [Organic order] The design is not rigidly planned, but is allowed to emerge over time in response to people's needs and activities. 
    \item [Participation] The users of a system should play a role in designing it---not just in providing feedback, but in shaping its direction.
    \item [Piecemeal growth] A campus needs to grow over time, like a living organism, little by little. It is better and also cheaper to build things in small projects rather than enormous buildings. 
    \item [Patterns] A shared language to guide improvement projects. This identifies a skeleton for what the university life and structures should be like. They sit at various scales, from big-picture to small. Examples include: open university (porous boundaries with the town), small spaces to exercise in (such as green space), and local transport (prioritizing pedestrians).
    \item [Diagnosis] Provides a framework for identifying areas of campus that violate each pattern (and thus are candidates for improvement). 
    \item [Coordination] Describes governance. Even the patterns themselves should be revisited and amended over time.
\end{description}

This particular master plan stood for some years and guided useful improvements around the University of Oregon's campus in that time. But eventually it was set aside, as universities have apparent priorities other than ensuring livelihood in their spaces \citep{Bryant1994}.

Though it was abandoned, it is clear that as a plan, Alexander's approach to university master planning was much better than the static map-of-the-future approach. While Alexander's work and principles here are certainly resonant with other currents in design over the decades, I would suggest that his work, particularly in \emph{The Oregon Experiment}, deserves renewed interest, particularly amidst current discussions of humanity-centered design. As I have written elsewhere \citep{Gorichanaz2023alexander}, though Alexander is best known in HCI for his work on design patterns, his broader life project has much relevance to HCI's goals of designing for humans and humanity (``humanity'' here in both senses of the word). 

\section{Toward a More Humble Humanity-Centered Design}

As discussed above, the current articulation of humanity-centered design includes paradoxes that make it run the risk of arrogance and eventually hubris. But hubristic humanity-centered design is not a foregone conclusion. As we strive to develop and operationalize humanity-centered design, humility is possible and preferred. Drawing lessons from Alexander's discussion of university master planning, here I will sketch some principles of a more humble humanity-centered design. 

In the spirit of humility, when it comes to human involvement, we must remember that a single universal solution or even set of priorities is not the best way forward. A single map of the future will not suffice. Perhaps, rather, we should strive to develop a diverse array of plans for the future, which are compatible where relevant. As an analogy, consider how the world's libraries do not rely on a single universal system for cataloging items (though at various moments in the past creating such a system was taken as a goal) but make use of union catalogs and other techniques for intercommunication across libraries. 

To put it generally, different people will care about different things and have different needs, and so pluralism must be supported. Following one of the principles from Alexander's master plan for the University of Oregon, organic order and participatory design should be the order of the day.

As a corollary, short-term increments should be considered preferable to long-term mapping. Whereas the hubristic approach to humanity-centered design presumes that we can discover (at least in principle) the best long-term solution to a large problem, a more humble approach suggests taking smaller steps to resolve the large problem, remaining agnostic about the form that the eventual solution may take and staying ever attentive to the problem itself. Alexander encapsulated this in his master plan with evolving governance, providing a meta-plan similar to how the U.S. Constitution describes procedures for its own amendment. (And consider how the original constitution runs about 4,000 words, while its 27 amendments run an additional 7,000 words.) Humanity-centered design work could follow suit. 

When it comes to participation, involvement from stakeholders (including indirect ones) is desirable, but at some point adding more participants has diminishing returns, followed by detriment (the too-many-cooks-in-the-kitchen phenomenon). Inclusive design should be encouraged but perhaps not pursued monomaniacally at the expense of other dimensions of the project. Further research may be able to help identify where those lines are, analogous to the way that research has helped the HCI field identify how many usability testing participants offer the best return on investment \citep{Nielsen2012b}. 

Finally, when it comes to ecological scope, things are more difficult. How to involve entities who cannot speak for themselves is a challenge, but the perfect should not be the enemy of the good. Even if nourishment is the only shared need we can identify between humans and, say, wild rice, that is enough to advocate for reasonable protections for wild rice's habitat, as has been seen in a recent Minnesota lawsuit \citep{Lovett2021}. That is, we can make a reasonable attempt to act as stewards of the world in which we find ourselves, and all the entities we find ourselves with, without presuming that we know every last detail so as to definitively engineer their well-being for all time. For now, a guide in this direction is provided in the work of Wakkary \citep{Wakkary2021}.

\section{Conclusion}

Humanity-centered design is a concept of emerging interest in HCI, one motivated by the limitations of human-centered design. Its overarching goal is certainly laudable: enabling a better world. Yet current articulations of humanity-centered design are incoherent in a number of ways, leading to questions of how exactly it can or should be implemented. 

In this piece, I have delineated four ways in which humanity-centered design is incoherent, chalking them all up to a tendency toward hubris. I have proposed that a more fruitful way forward would be a humble approach to humanity-centered design. Rather than a contradiction in terms, humility here refers to an organic, piecemeal, patterns-based approach to design that will be good for our being on this earth. And this may be felicitous indeed, seeing as the words ``humility'' and ``humanity'' both come the same etymological root meaning ``earth.''

\begin{acks}

Portions of the arguments presented here were originally published in my email newsletter (\url{https://ports.substack.com}). See the articles ``From Human- to Humanity-Centered Design'' (August 25, 2023) and ``Humanity-Centered Design without Hubris'' (September 1, 2023).

\end{acks}

\bibliographystyle{ACM-Reference-Format}
\bibliography{refs}

\end{document}